\documentclass[12pt,a4paper]{article}
\usepackage[utf8]{inputenc}
\usepackage[english]{babel}
\usepackage{amsmath}
\usepackage{amsfonts}
\usepackage{amssymb}
\usepackage{graphicx}
\usepackage{lmodern}
\usepackage[left=2.5cm,right=2.5cm,top=3cm,bottom=3cm]{geometry}
\usepackage{authblk}
\usepackage[hidelinks]{hyperref}
\usepackage[hyphenbreaks]{breakurl}
\usepackage[square,numbers,super,compress]{natbib}

\usepackage{lmodern}
\usepackage[T1]{fontenc}
\usepackage{color}

\usepackage{booktabs}

\title{\vspace{-2cm}Sample size re-estimation incorporating prior information on a nuisance parameter}	

\author[a]{Tobias M{\"u}tze\thanks{Correspondence: Tobias M{\"u}tze, Institut f{\"u}r Medizinische Statistik, Humboldtallee 32, 37073 G{\"o}ttingen, Germany. Email: tobias.muetze@med.uni-goettingen.de}}
\author[b]{Heinz Schmidli}
\author[a,c]{Tim Friede}
\affil[a]{Department of Medical Statistics, University Medical Center G{\"o}ttingen, G{\"o}ttingen, Germany}
\affil[b]{Statistical Methodology, Novartis Pharma AG, Basel, Switzerland}
\affil[c]{DZHK (German Centre for Cardiovascular Research), partner site G{\"o}ttingen, G{\"o}ttingen, Germany}

\providecommand{\keywords}[1]{\textit{Keywords:} #1}
\providecommand{\refas}[1]{
\vspace{0.5cm}
\noindent
\hspace{\dimexpr-\fboxrule-\fboxsep\relax}\fbox{%
  \begin{minipage}[t]{\linewidth}
   \textit{Reference as:}\\ #1
  \end{minipage}%
}
}

\date{\vspace{-5ex}}

\usepackage{fancyhdr}
\fancypagestyle{plain}{%
  \fancyhf{}
  
  \fancyfoot[L]{\textit{Post-print}}
  \fancyfoot[C]{\thepage}
}  
  
\begin{document}
\maketitle

\noindent\makebox[\width]{\rule{\textwidth}{0.2pt}}

\begin{abstract}
Prior information is often incorporated informally when planning a clinical trial.
Here, we present an approach on how to incorporate prior information, such as data from historical clinical trials, into the nuisance parameter based sample size re-estimation in a design with an internal pilot study.
We focus on trials with continuous endpoints in which the outcome variance is the nuisance parameter. 
For planning and analyzing the trial frequentist methods are considered. 
Moreover, the external information on the variance is summarized by the Bayesian meta-analytic-predictive (MAP) approach.
To incorporate external information into the sample size re-estimation, we propose to update the MAP prior based on the results of the internal pilot study and to re-estimate the sample size using an estimator from the posterior. 
By means of a simulation study, we compare the operating characteristics such as power and sample size distribution of the proposed procedure with the traditional sample size re-estimation approach which uses the pooled variance estimator.
The simulation study shows that, if no prior-data conflict is present, incorporating external information into the sample size re-estimation improves the operating characteristics  compared to the traditional approach. 
In the case of a prior-data conflict, that is when the variance of the ongoing clinical trial is unequal to the prior location, the performance of the traditional sample size re-estimation procedure is in general superior, even when the prior information is robustified. 
When considering to include prior information in sample size re-estimation, the potential gains should be balanced against the risks.
\end{abstract}

\noindent\makebox[\linewidth]{\rule{\textwidth}{0.2pt}}

\refas{
T. M{\"u}tze, H. Schmidli, T. Friede (2018). 
Sample size re-estimation incorporating prior information on a nuisance parameter.
Pharmaceutical Statistics, 17(2), 126-143.} 

\section{Introduction}
%
In clinical trials with a continuous outcome variable, the sample size planning is often based on the assumption of normally distributed data. 
In this case, the sample size is affected by the outcome variance and the assumed treatment effect size as well as the type I and type II error rates.
When planning the sample size, the treatment effect size is generally determined based on clinical relevancy and the type I and type II error rates are selected from a set of agreed upon values.
However, the outcome variance is usually unknown and determining it to calculate the sample size has been discussed extensively in literature.
If historical trials with designs similar to the planned trial are available, the variance for the sample size formula can be estimated from the historical trials. \cite{charpentier2016meta, pigott2008methodological, schmidli2016meta, julious2006sample}
Another possibility to obtain an estimate of the outcome variance is from a pilot study.
Pilot studies can either be internal or external, the difference being that internal pilot studies are part of the main clinical trial and external pilot studies are a separate entity. \cite{wittes1990role}
The variance estimate or a confidence limit of the variance obtained from an external pilot study can be considered for the sample size planning of a clinical trial. \cite{browne1995use, kieser1996use, whitehead2016estimating} 
However, there is no guarantee that the outcome variance of a clinical trial corresponds to the estimate obtained from an external source.
For situations in which information on the variance is uncertain, it has been proposed to include an internal pilot study into the design of a clinical trial and to re-estimate the outcome variance (or nuisance parameters in general), and thus the sample size of the ongoing clinical trial, based on the results of the internal pilot study.\cite{wittes1990role}
Nuisance parameter based sample size re-estimation in clinical trials has been studied extensively and we refer to reviews for a detailed recapitulation. \cite{friede2006sample, chuang2006sample, proschan2009sample, pritchett2015sample} 
It is worth emphasizing that nuisance parameter based sample size re-estimation can be performed either blinded or unblinded, in contrast to effect based sample size re-estimation, which is always performed unblinded. 
For a comparison of effect based versus nuisance parameter based sample size re-estimation and the respective operational and regulatory challenges, we refer to existing literature. \cite{pritchett2015sample, he2017addressing}  \\ \indent
In this manuscript we focus on two-arm parallel group superiority trials with normally distributed endpoints planned and analyzed using frequentist methods. 
We assume that the clinical trial design includes an internal pilot study with a nuisance parameter, that is the outcome variance, based sample size re-estimation.
Moreover, we also assume that prior information on the outcome variance is available as a meta-analytic-predictive (MAP) prior. \cite{schmidli2016meta}
We study several methods to incorporate prior information on the outcome variance into the  unblinded sample size re-estimation and assess the power and final sample size distribution of the respective sample size re-estimation procedures in a simulation study.
We focus on re-estimating the sample size such that the clinical trials maintains a prespecified power.
Our motivation for primarily focusing on incorporating information into the sample size re-estimation based on unblinded data instead of blinded data is of a computational nature. 
We then elaborate on why the findings for unblinded data are qualitatively the same as for incorporating information into the sample size re-estimation based on blinded data.
Incorporating prior information into the sample size re-estimation has already been proposed for binomial data in the early 1990s but not studied intensively. \cite{Gould1992}
More recently, Hartley introduced an approach to blinded sample size re-estimation for normal data. \cite{hartley2012adaptive} \\ \indent
This manuscript is structured as follows.
In Section \ref{sec:method} we elaborate the statistical model and recapitulate MAP priors as well as how to incorporate external information into the sample size planning.
In Section \ref{sec:SSR} we propose several procedures for incorporating prior information into the sample size re-estimation. 
Clinical trial examples are discussed in Section \ref{sec:example}. 
The performance characteristics of the introduced sample size re-estimation procedures incorporating prior information are studied in Section \ref{sec:simulation}. 
In Section \ref{sec:blinded} we discuss how prior information can be incorporated into the sample size re-estimation with blinded data and why the performance of the resulting procedures are qualitatively identical to the performance of the procedures based on unblinded data.
We conclude with a discussion in Section \ref{sec:discussion}.
\section{Statistical model and meta-analytic-predictive priors}
\label{sec:method}
This section is split into three parts. 
In the first part we outline the statistical model considered in this manuscript and the classical frequentist approach of sample size planning. 
In the second part we recapitulate MAP priors.
In the third part several approaches for incorporating prior information into frequentist sample size planning are discussed. 
\subsection{Statistical model}
Here we consider a two-arm parallel group superiority trial with normally distributed endpoints. 
More precisely, let $X_{ij}$ be the random variable modeling observation $j=1,\ldots, n_{i}$ in group $i=T,C$. 
Here, $i=T$ indicates the treatment group and $i=C$ the control group.
The randomization ratio is given by $k=n_{C}/n_{T}$ and the total sample size is given by $n=n_{T} + n_{C}$.
The random variables are independently normally distributed given the group mean $\mu_i$ and the variance $\sigma^2$, i.e.
\begin{align*}
X_{ij}|\mu_{i}, \sigma^2 \sim \mathcal{N}\left(\mu_i, \sigma^2 \right).
\end{align*}
Larger values of $\mu_i,\, i=T,C,$ are considered to be more desirable.
We focus on the frequentist hypothesis testing problem
\begin{align*}
H_{0}: \mu_T \leq \mu_C \qquad \text{vs.} \qquad H_{1}: \mu_T > \mu_C.
\end{align*}
The most common test for the hypothesis $H_0$ is the two-sample Student's \textit{t}-test with test statistic
\begin{align*}
T = \frac{\bar{X}_{T} - \bar{X}_{C}}{\hat{\sigma}\sqrt{\frac{1}{n_{T}}+\frac{1}{n_{C}}}}.
\end{align*}
The sample standard deviation $\hat{\sigma}$ is the square root of the pooled sample variance.
The null hypothesis $H_0$ can be rejected if the test statistic $T$ exceeds $t_{n-2, 1-\alpha}$, the $(1-\alpha)$-quantile of a \textit{t}-distribution with $n-2$ degrees of freedom.
Under the alternative hypothesis $H_{1}$, the test statistic $T$ follows a noncentral \textit{t}-distribution with $n-2$ degrees of freedom and noncentrality parameter 
\begin{align*}
\lambda = \frac{\delta}{\sigma\sqrt{\frac{1}{n_{T}}+\frac{1}{n_{C}}}}
= \frac{\sqrt{nk}}{k+1}\frac{\delta}{\sigma}.
\end{align*}
Thus, the power of Student's \textit{t}-test can be expressed as a function $B(n, \sigma^2, \delta, k)$ of the sample size $n$, the variance $\sigma^2$, the effect size $\delta$, and the ratio $k=n_{C}/n_{T}$, i.e.
\begin{align*}
B(n, \sigma^2, \delta, k) 
= \mathbb{P}\left(T\geq t_{n - 2, 1-\alpha}\right)
= 1 - F_{nct}\left(t_{n - 2, 1-\alpha};\lambda, n-2\right)
\end{align*}
with $F_{nct}\left(\cdot\,;\lambda, \nu\right)$ the cumulative distribution function of a noncentral \textit{t}-distribution with noncentrality parameter $\lambda$ and $\nu$ degrees of freedom.
The total sample size $n$ to test the hypothesis $H_0$ with a prespecified nominal power of $1-\beta$ is given by 
\begin{align}
\label{eq:SSttest}
n = \min \left\{ \tilde{n} \in \mathbb{N}: F_{nct}\left(t_{\tilde{n} - 2, 1-\alpha};\lambda^{*}, \tilde{n}-2\right)\leq \beta;\quad \lambda^{*}=\sqrt{\tilde{n}k}\frac{\delta^{*}}{\sigma(k+1)} \right\}.
\end{align}
Here, $\delta^{*} > 0$ is the assumed effect size under the alternative hypothesis.
Especially for large sample sizes, closed form approximations of the above sample size formula based on normal quantiles or quantiles of Student's \textit{t}-distribution exist and are commonly applied in practice. 
\subsection{Meta-analytic-predictive priors}
\label{sec:MAP}
Planning the sample size of a trial based on \eqref{eq:SSttest} requires knowledge about the variance $\sigma^2$. 
In practice, information about the variance $\sigma^2$ is often gathered from historical studies.  
Schmidli et al. formalized information gathering for nuisance parameters based on the meta-analytic-predictive (MAP) approach. \cite{schmidli2016meta}
In the following, we give a brief introduction to the MAP approach for variances. 
The idea is to perform a random effects meta-analysis using a normal Bayesian hierarchical model for the logarithm of the variance. 
The resulting posterior predictive distribution for the variance is the MAP prior which is used to predict the variance of a new clinical trial. 
In detail, let $j=1,\ldots,J$ be the index for $J$ historical clinical trials and  let  $\hat{\sigma}^2_{j}$  be the  sample variance and $\nu_j$ the respective degrees of freedom.
The unknown true variance of trial $j=1,\ldots,J$ is denoted by $\sigma_{j}^{2}$.
Considering that we are focusing on clinical trials with normal data, we assume that the sample variances follow a $\chi^2$-distribution in the sense
\begin{align*}
\frac{\nu_j}{\sigma_j^2}\hat{\sigma}_{j}^{2} \Big| \sigma_j^2 \sim \chi_{\nu_j}^2.
\end{align*}
We note that the $\chi^2$-distribution is a special case of the Gamma distribution and it follows that
\begin{align*}
\hat{\sigma}^2_{j} \big| \sigma_j^2 \sim  
\operatorname{Gamma}\left(0.5\nu_j, 0.5\frac{\nu_j}{\sigma_j^2} \right).
\end{align*}
A random variable following a Gamma distribution with shape parameter $a$ and rate parameter $b$, $\operatorname{Gamma}(a,b)$, has mean $a/b$ and variance $a/b^2$.
To gather information about the variance $\sigma^2_{new}$ of a new, to be planned clinical trial from the variances of historical clinical trial, the variance $\sigma^2_{new}$ and the variances $\sigma^2_{j}$ of historical clinical trials have to be linked.
As it is common in random effects meta-analyses, we assume that the variances origin from the same distribution.
Here, we assume that the log-transformed variances $\theta_{new}=\log(\sigma^2_{new}), \theta_{1}=\log(\sigma^2_{1}), \ldots, \theta_{J}=\log(\sigma^2_{J})$ are independent and identically normally distributed,
\begin{align*}
\theta_{new}, \theta_{1}, \ldots, \theta_{J} \sim 
\mathcal{N}\left(\mu, \tau^2\right).
\end{align*}
The random effects meta-analysis can be performed using a Bayesian hierarchical model.
Thereto, prior distributions for the mean $\mu$ and the between-trial standard deviation $\tau$ have to be selected.
Common choices are weakly informed priors such as a normal distribution with a large variance for the mean and a half-normal distribution for the standard deviation.\cite{schmidli2016meta}
The posterior density $p(\mu,\tau,\theta_{new}|\hat{\sigma}_{1}^{2}, \ldots, \hat{\sigma}_{J}^{2})$ cannot be calculated analytically. 
However, through Markov chain Monte Carlo (MCMC) computations random samples of the parameter vector $(\mu, \tau, \theta_{new})$ can be generated. 
The random numbers for $\theta_{new}$ can then be transformed to obtain random numbers for the posterior predictive distribution of the variance of a new trial $\sigma^{2}_{new}$.
The posterior predictive distribution of $\sigma^{2}_{new}$ is the MAP prior for the variance.
The MAP prior for the variance generally does not have a closed form expression but, just like every prior distribution, it can be approximated by a mixture of conjugated priors. \cite{dalal1983approximating}
For the sake of convenience, the prior of the precision $\omega_{new} = 1/\sigma_{new}^2$, and not the prior of the variance $\sigma_{new}^2$, is approximated by a mixture of conjugated priors.
For a normal model, the conjugated prior of the precision is the Gamma distribution.
Therefore, in this manuscript we assume that the prior distribution for the precision $\omega_{new} = 1/\sigma^2_{new}$ is a mixture of Gamma distributions
\begin{align*}
\omega_{new} \sim \sum_{l=1}^{L}w_{l}\,\operatorname{Gamma}(a_{l}, b_{l}).
\end{align*}
When approximating the MAP prior, the parameters $w_{l}, a_{l}$, and $b_{l}$ are obtained by calculating the maximum-likelihood estimators from the MCMC random sample of the transformation $1/\exp(\theta_{new})$.
The number of mixture components $L$ is chosen such that the resulting fitted mixture distribution approximates the MCMC random sample with the desired precision. 
No closed form expression for the maximum-likelihood estimators exists, but literature on their numerical calculation is available. \cite{almhana2006}
For details on the methodological background of the MAP approach and its use to summarize historical information on the variance, we refer to Schmidli et al. \cite{schmidli2016meta, schmidli2014robust} 
Since $\omega_{new}$ follows a mixture of Gamma distributions, the variance $\sigma^2_{new}$ follows a mixture of inverse Gamma distributions with the same weights, shape parameters, and rate parameters. 
An inverse Gamma distributed random variable with shape parameter $a$ and rate parameter $b$, $\operatorname{InvGamma}(a,b)$, has mean $b/(a-1)$ and variance $b^2/((a-1)^2(a-2))$.\\ \indent
The MAP prior for the variance might mismatch the variance observed in a new clinical trial, i.e. a prior-data conflict can be present.
More precisely, a prior-data conflict can for example be defined as the case when the observed variance is outside a $95\%$ probability interval of the prior-predictive distribution, similar to Box (1980). \cite{box1980}
Schmidli et al. introduced a robustified version of MAP priors which aims to mitigate the risk of a prior-data conflict.\cite{schmidli2014robust}
We briefly recapitulate their robustification of MAP priors. 
Thereto, let $p_{MAP}(\cdot)$ be the MAP prior obtained from historical data and let  $p_{V}(\cdot)$ be a vague conjugate prior.
Moreover, let $w_{R}$ be the prior probability of a prior-data conflict.
Then, a robustified MAP prior with density $p_{rMAP}(\cdot)$ is the mixture distribution of the MAP prior and the vague conjugate prior with mixture probability $w_{R}$, that is
\begin{align*}
p_{rMAP}(x) = w_{R}\,p_{V}(x) + (1-w_{R})p_{MAP}(x).
\end{align*}
The choice of the prior probability $w_{R}$ of a prior-data conflict reflects the initial information of how likely a prior-data conflict is.
In this sense, the choice of $w_{R}$ has to be made on a case-to-case basis based on the available information and, for instance, on the desired operating characteristics of the resulting robustified prior. 
An example in which the prior information for the nuisance parameter was robustified during the sample size planning of a clinical trial was discussed by Schmidli et al.\cite{schmidli2016meta}
Moreover, it is also important to emphasize that the prior probability $w_{R}$ of a prior-data conflict gets updated when the posterior distribution is calculated. 
\subsection{Sample size planning based on prior information on the nuisance parameter}
In the following we outline how prior information on the nuisance parameter can be incorporated into the frequentist sample size calculation of a clinical trial. 
The methods for including prior information on the nuisance parameter into the sample size re-estimation, which we will present in the next section, follow the same principles.
To plan the sample size $n$ of a new trial based on the MAP prior $p_{\sigma^2}(\cdot)$ of the variance $\sigma^2$, a Bayes estimator of $\sigma^2$ can be plugged into Formula \eqref{eq:SSttest}. \cite{schmidli2016meta}
There are various ways of defining a Bayes estimator; in this sense, Bayes estimators are not unique. \cite{mood1974} 
Here, we focus on the mean $\hat{\sigma}^2_{mean}$ and the median $\hat{\sigma}^2_{med}$ of the MAP prior which minimize the squared error risk and the absolute deviation, respectively.
The sample size can also be planned based on prior information on the standard deviation $\sigma$ or on the precision $\omega=1/\sigma^2$. 
However, if the Bayes estimator is not transformation invariant, the resulting sample size will differ from the sample size based on prior information on the variance.
When the sample size is determined based on a location parameter of the MAP prior, the uncertainty of the prior information is not considered in the planning process. 
Alternative approaches which consider the variability of the prior information can either choose percentiles of the MAP prior in Equation \eqref{eq:SSttest} or base the sample size on the unconditional power. 
The unconditional power concerning $\sigma^2$ is the function 
\begin{align*}
\tilde{B}(n, \delta, k) = \int B(n, x, \delta, k)p_{\sigma^2}(x) \mathrm{d}x.
\end{align*}
The total sample size $n$ for testing the hypothesis $H_0$ which is then defined by 
\begin{align}
\label{eq:SSttest2}
n = \min \left\{ \tilde{n} \in \mathbb{N}: \tilde{B}(\tilde{n}, \delta^{*}, k)\geq 1-\beta \right\}.
\end{align}
It is worth emphasizing that the sample size obtained from the unconditional power generally differs from the sample size obtained with Formula \eqref{eq:SSttest}.
\section{Nuisance parameter based sample size re-estima\-tion using MAP priors}
\label{sec:SSR}
In this section we briefly summarize the general idea of adjusting the sample size based on an estimate of the nuisance parameter in designs with internal pilot study and then propose methods which incorporate prior information on the variance in the form of MAP priors into the sample size re-estimation.
As the name implies, in nuisance parameter based sample size re-estimation, the sample size is altered after the internal pilot study based on an estimate of the nuisance parameter.
The estimation of the nuisance parameter can either be done based on blinded or unblinded data. 
According Section 4.4 of ICH guideline E9, it must be addressed whether and how the blindness during the sample size re-estimation was maintained. \cite{ICHE9}
Other regulatory publications recommend to perform the nuisance parameter based sample size re-estimation blinded whenever possible, such as a reflection paper by the Committee for Medicinal Products for Human Use (CHMP). \cite{CHMPadapt}
The recommended method for nuisance parameter based sample size re-estimation in two-arm parallel group trials with continuous data is the blinded one-sample variance estimator which estimates the unknown variance by the sample variance of the blinded data. \cite{phillips2006adaptive, gallo2006adaptive, friede2013blinded}
The one-sample variance estimator results in a sample size re-estimation procedure which meets the power and controls the type I error rate for practically relevant internal pilot study sizes.
Although the one-sample variance estimator is the recommended method for nuisance parameter based sample size re-estimation, we first introduce the sample size re-estimation incorporating prior information for unblinded data. 
The primary reason is that unblinding of the internal pilot study leads to closed form expressions for the sample size re-estimation incorporating external information.
We will show in Section \ref{sec:blinded} that the performance of the sample size re-estimation procedure incorporation prior information based on blinded data is qualitatively the same as for the procedure relying on unblinded data.
Altering the sample size of an ongoing clinical trial based on unblinded data from an internal pilot study has first been proposed by Wittes et al.\cite{wittes1990role}
In detail, the idea is to perform an internal pilot study of size $n_1$, with $n_{1T}$ and $n_{1C}$ denoting the group specific sample sizes within the internal pilot study.
After the outcome measure of the $n_1$ patients is obtained, the pooled variances $\hat{\sigma}^{2}_{1,pool}$ from the unblinded data is calculated.
The sample size  is then re-estimated based on the estimate $\hat{\sigma}^{2}_{1,pool}$ using Equation \eqref{eq:SSttest}.
Let $\hat{n}_{reest}$ denote the re-estimated sample size.
Here, we consider the final sample size of the clinical trial to be the maximum of the re-estimated sample size and the internal pilot study sample size, i.e. $\hat{n}_{final} = \max\{\hat{n}_{reest}, n_{1}\}$. \cite{Birkett1994}
Alternatively, the final sample size can be set to be the maximum of the initially planned sample size and the re-estimated sample size in which case  the sample size re-estimation would not be able to reduce the initially set sample size. \cite{wittes1990role}
In the case of constrained resources it can also be reasonable to additionally include an upper limit for the final sample size. \cite{Gould1992}
Sample size re-estimation based on the pooled sample variance is able to adjust for a misspecified variance during the planning of the clinical trial if the internal pilot study is reasonably sized. \cite{wittes1990role, Birkett1994, friede2001comparison}
This sample size re-estimation approach inflates the type I error rate and due to the unblinding might trigger the need of an Independent Data Monitoring Committee (IDMC) to preserve the blindness of the study personnel. 
It is worth mentioning that the type I error rate inflation can be minimized when a bias correction is applied to the variance estimator in the test statistic at the end of the trial. \cite{miller2005variance}
We do not apply this bias correction here because it qualitatively does not change the performance of the sample size re-estimation procedure incorporating prior information.\\ \indent
When the sample size of a clinical trial is planned utilizing prior information or in general when prior information on the nuisance parameter is available, it seems intuitive to incorporate said prior information into the sample size re-estimation. 
We propose to incorporate prior information on the nuisance parameter into the sample size re-estimation by updating the prior using the data from the internal pilot study and then re-calculating the sample size with \eqref{eq:SSttest} based on a Bayes estimator of the variance obtained from the posterior distribution. 
In more detail, let $X_{ij}$ be the random variable modeling observation $j=1,\ldots, n_{1i}$ in group $i=T,C$ after the internal pilot study and let $\omega = 1/\sigma^2$ be the precision. 
The prior information on the parameters $(\mu_T, \mu_C, \omega)$ is characterized by the prior density $p(\mu_T, \mu_C, \omega)$. 
Thus, for the posterior density after the internal pilot study holds
\begin{align*}
p\left(\mu_T, \mu_C, \omega|\bar{X}_{1T}, \bar{X}_{1C}, \hat{\sigma}^2_{1,pool}\right) \propto p\left(\bar{X}_{1T}, \bar{X}_{1C}, \hat{\sigma}^2_{1,pool}|\mu_T, \mu_C, \omega\right) p(\mu_T, \mu_C, \omega).
\end{align*}
Here, $\bar{X}_{1T}$, $\bar{X}_{1C}$, and $\hat{\sigma}^2_{1,pool}$ denote the sample mean in the treatment group, the sample mean in the control group, and the pooled sample variance, respectively, obtained from the unblinded data of the internal pilot study.
In this manuscript we focus on the specific case of an improper uniform prior for the means which is a prior independent of the prior for the precision,
\begin{align*}
&p(\mu_T, \mu_C) = p(\mu_T) p(\mu_C) = 1,\\
&p(\mu_T, \mu_C, \omega) = p(\mu_T, \mu_C) p(\omega).
\end{align*}
As mentioned in the previous section, we assume that the precision $\omega$ has a mixture of Gamma distributions as the prior. 
In the assumed model, this prior distribution is a conjugate prior and thus the marginal posterior for the precision is again a mixture of Gamma distributions, \cite{bernardo1994bayesian}
\begin{align*}
\omega \big| \bar{X}_{1T}, \bar{X}_{1C}, \hat{\sigma}^2_{1,pool} \sim \sum_{l=1}^{L} w_{l}^{*} \operatorname{Gamma}\left(a_l + \frac{n_1-2}{2}, b_l + \frac{n_{1}-2}{2} \hat{\sigma}^2_{1,pool}\right).
\end{align*}
The updated weights $w_{l}^{*}$ are given by
\begin{align*}
w_{l}^{*} = \frac{r_l}{\sum_{l=1}^{L}r_l}
\end{align*}
where the single components $r_l$ are calculated by
\begin{align*}
r_{l} = w_{l} \frac{\Gamma\left(a_{l} + 0.5(n_{1}-2)\right)}{\left(b_{l} + 0.5(n_{1}-2)\hat{\sigma}^2_{1,pool}\right)^{a_{l}+0.5(n_1-2)}}
\frac{b_{l}^{a_{l}}}{\Gamma(a_{l})}.
\end{align*}
Thus, the posterior for the variance $\sigma^2$ follows a mixture of inverse Gamma distributions.
We denote the mean and the median of the posterior distribution of $\sigma^2$ as $\hat{\sigma}^{2}_{1,mean}$ and $\hat{\sigma}^{2}_{1,med}$, respectively.
The mean $\hat{\sigma}^{2}_{1,mean}$ of the posterior distribution is the weighted mean of inverse Gamma distributions,
\begin{align*}
\hat{\sigma}^{2}_{1,mean} = \sum_{l=1}^{L}w_{l}^{*} \frac{b_{l} + \hat{\sigma}_{1,pool}^{2} (n_{1}-2)/2}{a_{l} + (n_{1}-2)/2-1}.
\end{align*}
The median of an inverse Gamma distribution and the median of a mixture of inverse Gamma distributions do not have closed form expressions and must be calculated iteratively.
The re-estimated sample size $\hat{n}_{reest}$ is obtained through  \eqref{eq:SSttest} by plugging in a Bayes estimator for the variance. 
Analogously to the planning of a clinical trial, the sample size can also be determined by means of the unconditional power, confer  \eqref{eq:SSttest2}, or by plugging in percentiles of the posterior distribution into \eqref{eq:SSttest}.
\section{Clinical trial examples}
\label{sec:example}
In this section we discuss two examples in which we gather information on the variance from historical clinical trials and predict the variance of a new trial using the meta-analytic-predictive approach. 
Then, we study how the prior information about the variance affects the re-estimated sample size when incorporated into the sample size re-estimation.
\subsection{St John's wort for major depression}
\begin{figure}[!htb]
 \centering
 \includegraphics[width=1\linewidth]{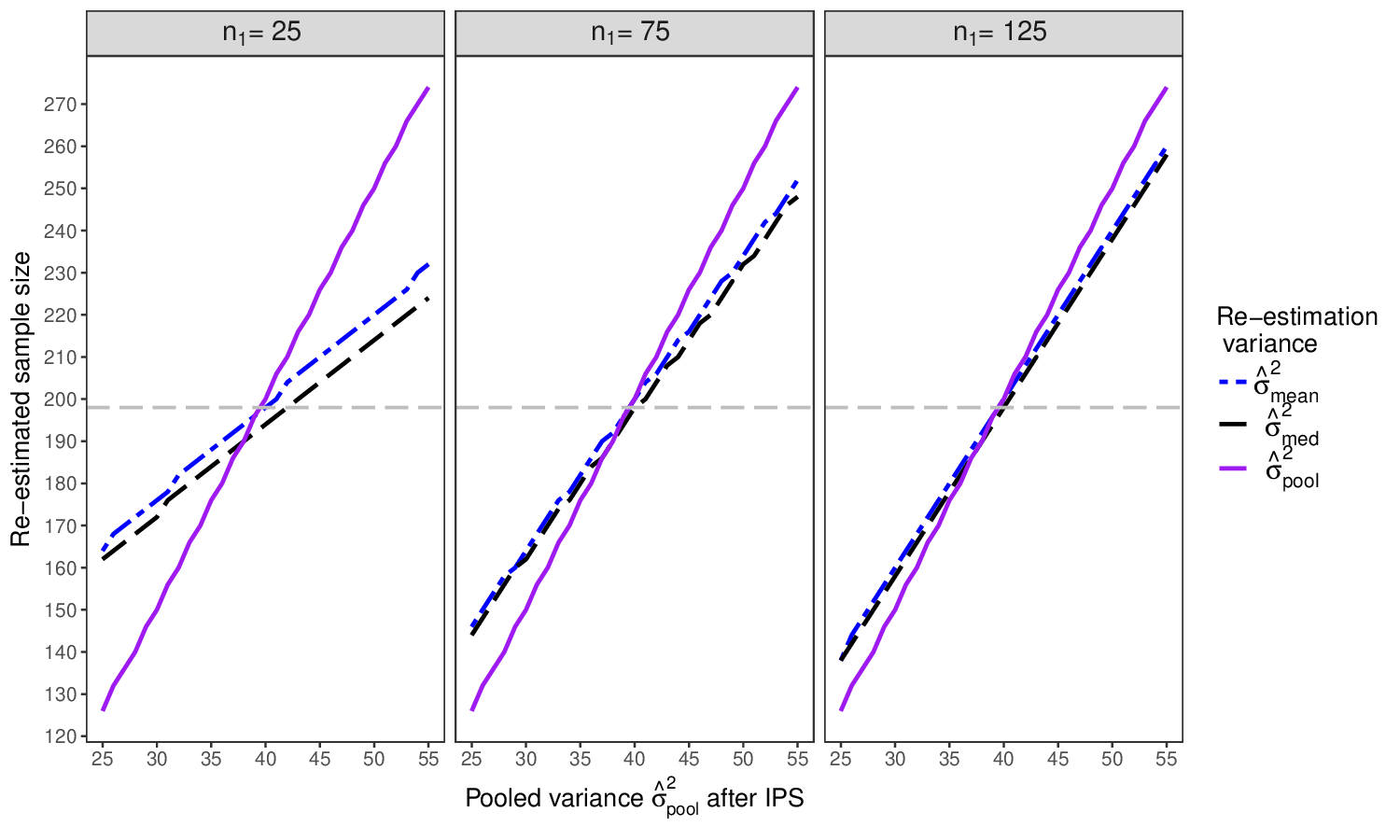}
  \caption{Re-estimated sample size for the unblinded sample size re-estimation based on the pooled variance $\hat{\sigma}^{2}_{pool}$ and for the sample size re-estimation incorporating prior information based on the posterior mean $\hat{\sigma}^{2}_{mean}$ and on the posterior median $\hat{\sigma}^{2}_{med}$, respectively.}
   \label{fig:ExampleHamd}
\end{figure}

The first example focuses on the use of St John's wort for major depression. 
Linde et al. published a meta-analysis summarizing the effects of St John's wort for major depression for a variety of endpoints. \cite{linde2008john}
Here, we focus on the Hamilton Rating Scale for Depression (HAMD) score after four weeks of treatment. \cite{hamilton1960rating}
In Analysis 2.3 of their review, Linde et al. summarize the results of eleven trials comparing hypericum (St John's wort) versus placebo. 
From the reported results we obtain a pooled variance estimator and the respective number of degrees of freedom for each trial.
The MAP prior for the precision and, thus, the variance is obtained  through a Bayesian meta-analysis of the sample variances as outlined in Section \ref{sec:MAP}. 
The MAP prior of the precision $\omega=1/\sigma^2$ can be approximated by the following Gamma mixture distribution
\begin{align*}
0.16 \operatorname{Gamma}(4.6, 140.4) + 
0.84 \operatorname{Gamma}(18.2, 689.3).
\end{align*}
The prior effective sample size can be calculated as the product of the degrees of freedom of the historical trials multiplied by the ratio of the predictive variance assuming no between-trial heterogeneity and the predictive variance accounted for between-trial heterogeneity. \cite{neuenschwander2010summarizing}
Here, this leads to an prior effective sample size of $ESS=24$.
Table \ref{table:stjohns} list characteristics of the MAP priors. 
\begin{table}[ht]
\caption{Summaries of the MAP priors for the variance $\sigma^2$, the standard deviation $\sigma$, and the precision $\omega$.}
\label{table:stjohns}
\begin{center}
\begin{tabular}{l*{1} {c}{c}{c}{c}{c}}
\toprule
\textbf{Parameter} & \textbf{Mean} & \textbf{SD} & \textbf{Median} & \textbf{$2.5\%$ quantile} & \textbf{$97.5\%$ quantile}  \\ \midrule
$\sigma^2$ & 39.56 & 12.56 & 37.93 & 21.11 & 68.52 \\
$\sigma$ & 6.22 & 0.93 & 6.16 & 4.59 & 8.27 \\
$\omega$ & 0.0276 & 0.0097 & 0.0267 & 0.0146 & 0.0474 \\
\bottomrule
\end{tabular}
\end{center}
\end{table}
Table \ref{table:stjohns} highlights that the prior information on the outcome variance is quite uncertain. 
In the following we study how the prior information affects the sample size when incorporated into the sample size re-estimation. 
Thereto, we assume internal pilot study sizes of $n_1=25, 75, 125$ and that the sample size re-estimation is performed with an assumed effect of $\delta^{*}=2.515$ which corresponds to a standardized effect of about $0.4$ for a variance of $\sigma^2=39.56$. 
For this parameter combination, the fixed design sample size would be $n=198$.
In Figure \ref{fig:ExampleHamd} the re-estimated sample size is plotted against the pooled sample variance $\hat{\sigma}^{2}_{pool}$ obtained from the internal pilot study.
Figure \ref{fig:ExampleHamd} shows that the re-estimated sample sizes based on the Bayes estimators, that are the posterior mean and the posterior median, are not increasing as steep as the re-estimated sample size based on the pooled sample variance.
Moreover, the difference between the re-estimated sample sizes with and without incorporated prior information decreases as the internal pilot study sample size $n_1$ increases.
In other words, the influence of the prior information on the re-estimated sample size decreases as the internal pilot study sample size increases.
\subsection{Interventions for controlling blood pressure}
To improve the blood pressure control of hypertensive patients a variety of interventions such as self-monitoring, patient education, health care provider education, appointment reminders, etc. have been proposed.
In a meta-analysis, Glynn et al. summarized the available literature on  the effect of various interventions on the blood pressure. \cite{glynn2010interventions}
In our example, we focus on the reported results about the systolic blood pressure for patients who self-monitored their blood pressure, confer Analysis 1.1 in Glynn et al. \cite{glynn2010interventions}
With the reported sample variances, we proceed as in the previous example: we calculate a pooled variance for each study and then perform an Bayesian meta-analysis for the sample variances. 
The mixture of Gamma distributions 
\begin{align*}
0.29 \operatorname{Gamma}(10.28, 2298.63) + 
0.71 \operatorname{Gamma}(38.46, 9366.28).
\end{align*}
approximates the MAP prior of the precision $\omega=1/\sigma^2$.
The effective sample size is $ESS=41$.
Summary statistics of the MAP priors of the variance, standard deviation, and the precision are listed in Table \ref{table:bp}.
\begin{table}[ht]
\caption{Summaries of the MAP priors for the variance $\sigma^2$, the standard deviation $\sigma$, and the precision $\omega$.}
\label{table:bp}
\begin{center}
\begin{tabular}{l*{1} {c}{c}{c}{c}{c}}
\toprule
\textbf{Parameter} & \textbf{Mean} & \textbf{SD} & \textbf{Median} & \textbf{$2.5\%$ quantile} & \textbf{$97.5\%$ quantile}  \\ \midrule
$\sigma^2$ & 251.47 & 58.24 & 244.7 & 157.8 & 385.7 \\
$\sigma$ & 15.76 & 1.76 & 15.64 & 12.56 & 19.64 \\
$\omega$ & 0.0042 & 0.00094 & 0.0041 & 0.0026 & 0.0063 \\
\bottomrule
\end{tabular}
\end{center}
\end{table}
Analogously to the first example, we study how the prior affects the re-estimated sample size. 
As before, the internal pilot study sample sizes are assumed to be $n_1=25, 75, 125$ and the assumed effect is again chosen such that a standardized effect of $0.4$ is obtained for the prior mean of the variance, hence $\delta^{*}=6.343$.
The sample size in a fixed sample design would be $n=198$.
\begin{figure}[!htb]
 \centering
 \includegraphics[width=1\linewidth]{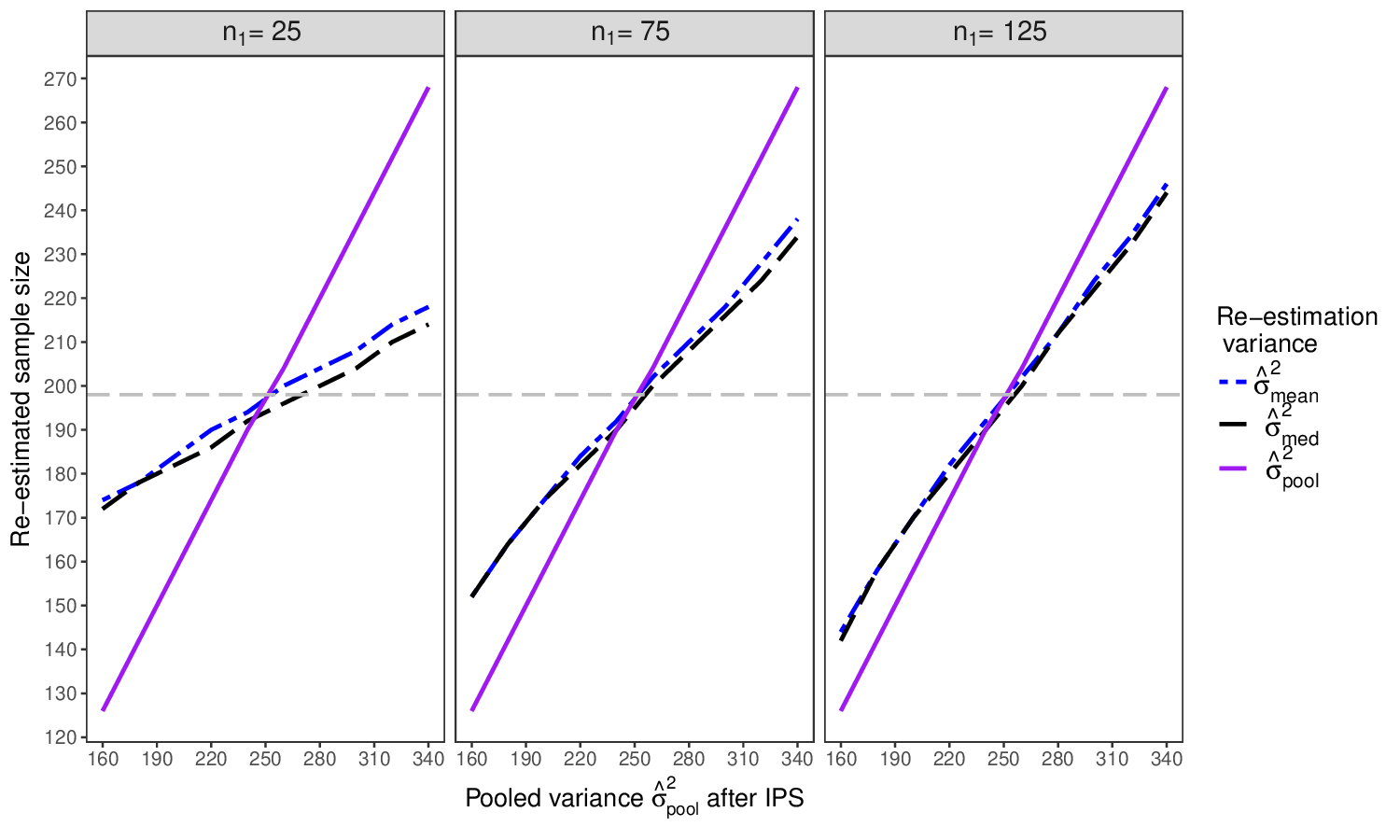}
  \caption{Re-estimated sample size for the unblinded sample size re-estimation based on the pooled variance $\hat{\sigma}^{2}_{pool}$ and for the sample size re-estimation incorporating prior information based on the posterior mean $\hat{\sigma}^{2}_{mean}$ and on the posterior median $\hat{\sigma}^{2}_{med}$, respectively.}
   \label{fig:ExampleSbp}
\end{figure}
The results shown in Figure \ref{fig:ExampleSbp} are qualitatively the same as in Figure \ref{fig:ExampleHamd}.
However, since the prior effective sample size  is larger in the second example, the difference between the re-estimated sample sizes from the methods with and without incorporated prior information is larger, too.
\section{Performance of the proposed sample size re-esti\-mation procedure}
\label{sec:simulation}
\begin{figure}[!htb]
 \centering
 \includegraphics[width=1\linewidth]{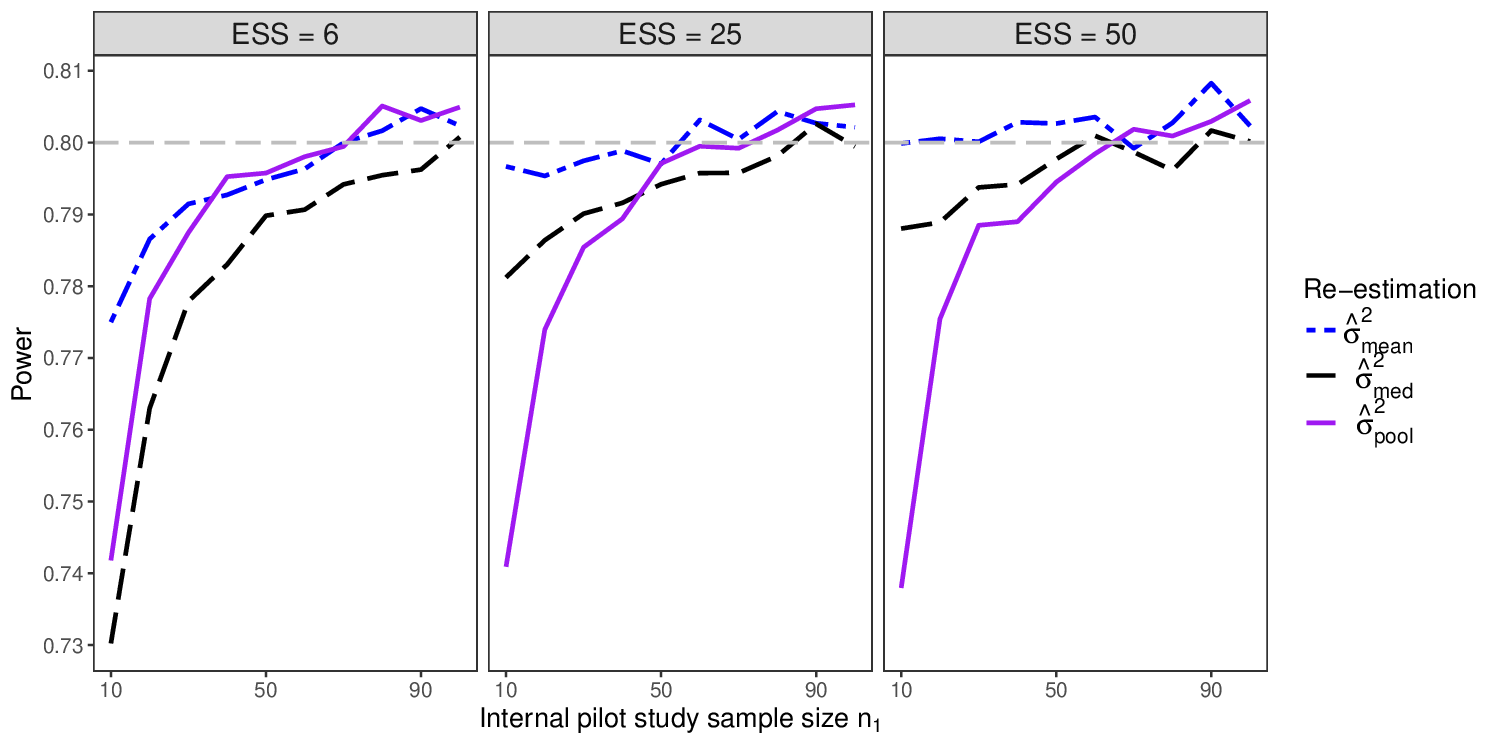}
  \caption{Power of the different sample size re-estimation procedures for prior effective sample sizes of $ESS=6,25,50$. The expected value of the prior information is identical to the true variance $\sigma^2=1$, thus, no prior-data conflict is present. The horizontal grey line depicts the nominal power of $80\%$.}
   \label{fig:Sce1power}
\end{figure}

\begin{figure}[htb]
 \centering
 \includegraphics[width=1\linewidth]{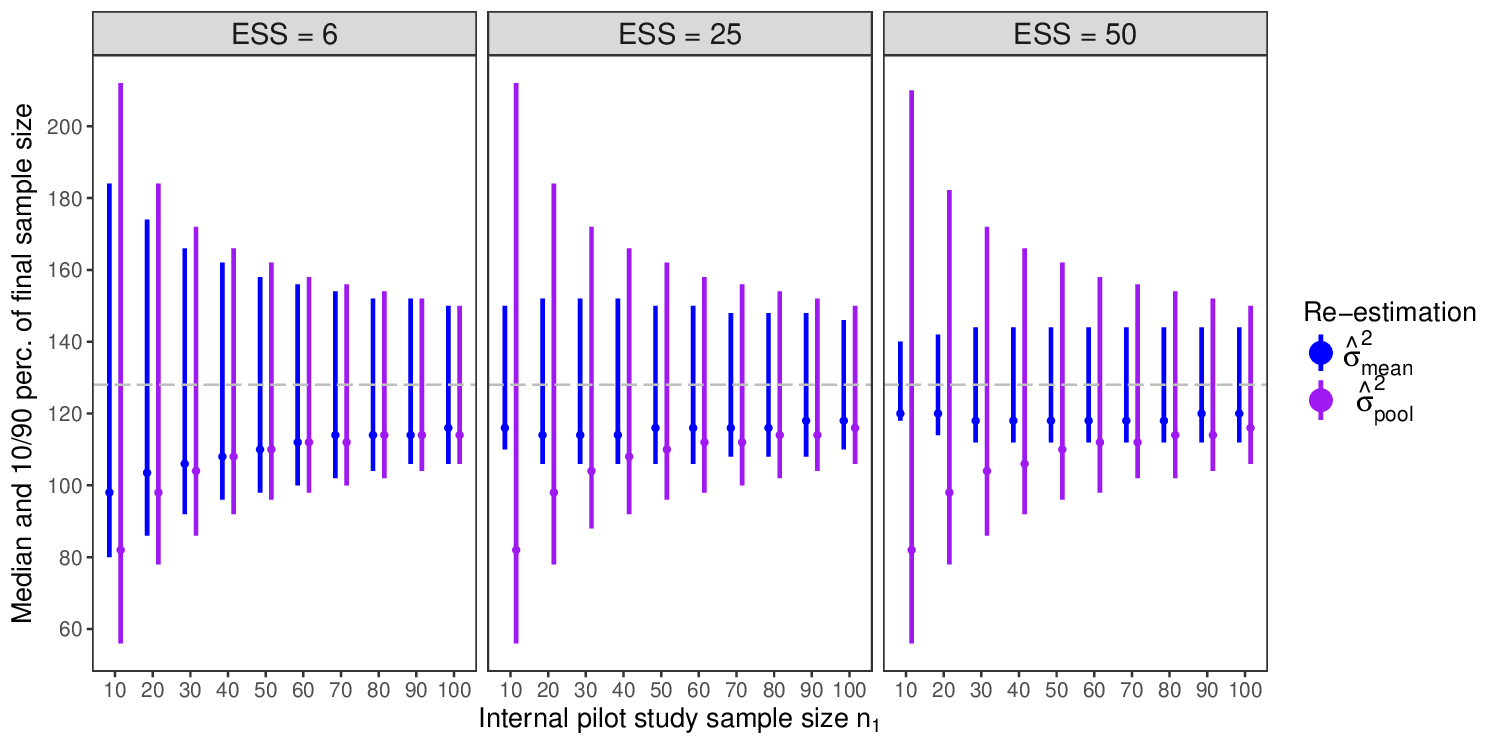}
  \caption{Median and percentiles ($10\%$ and $90\%$) of the final sample size. The expected value of the prior information is identical to the true variance $\sigma^2=1$, thus, no prior-data conflict is present. The horizontal grey line depicts the fixed design sample size of $n=128$.}
 \label{fig:Sce1ss}
\end{figure}

\begin{figure}[htb]
    \centering
    \includegraphics[width=1\linewidth]{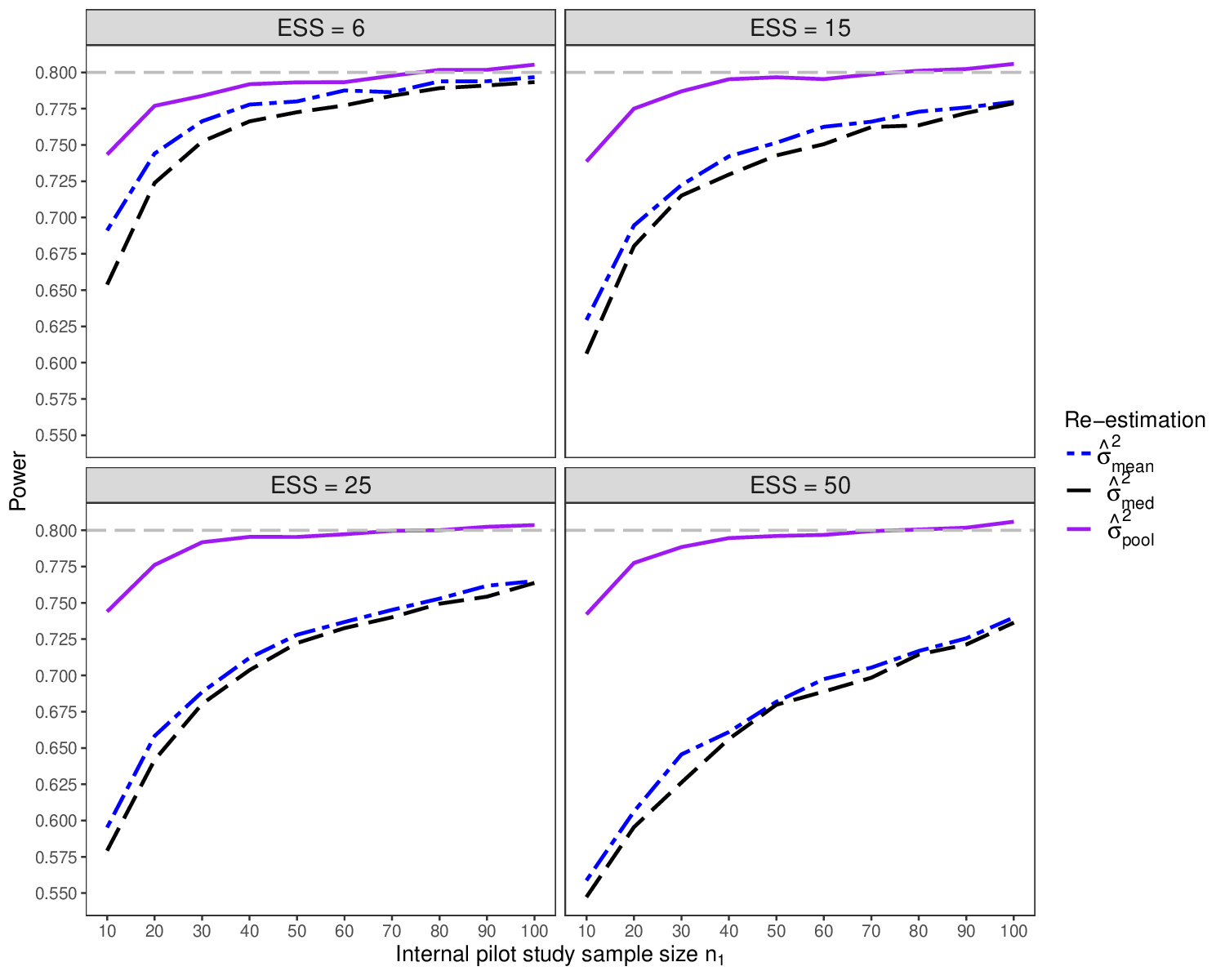}
    \caption{Power of the different sample size re-estimation procedures for prior effective sample sizes of $ESS=6,15, 25,50$. 
    The prior is in conflict with the data in the sense that the prior has an expected value of $0.49$ while the true variance is $\sigma^2=1$. The horizontal grey line depicts the nominal power of $80\%$.}
    \label{fig:Sec2power}
\end{figure}

\begin{figure}[htb]
 \centering
 \includegraphics[width=1\linewidth]{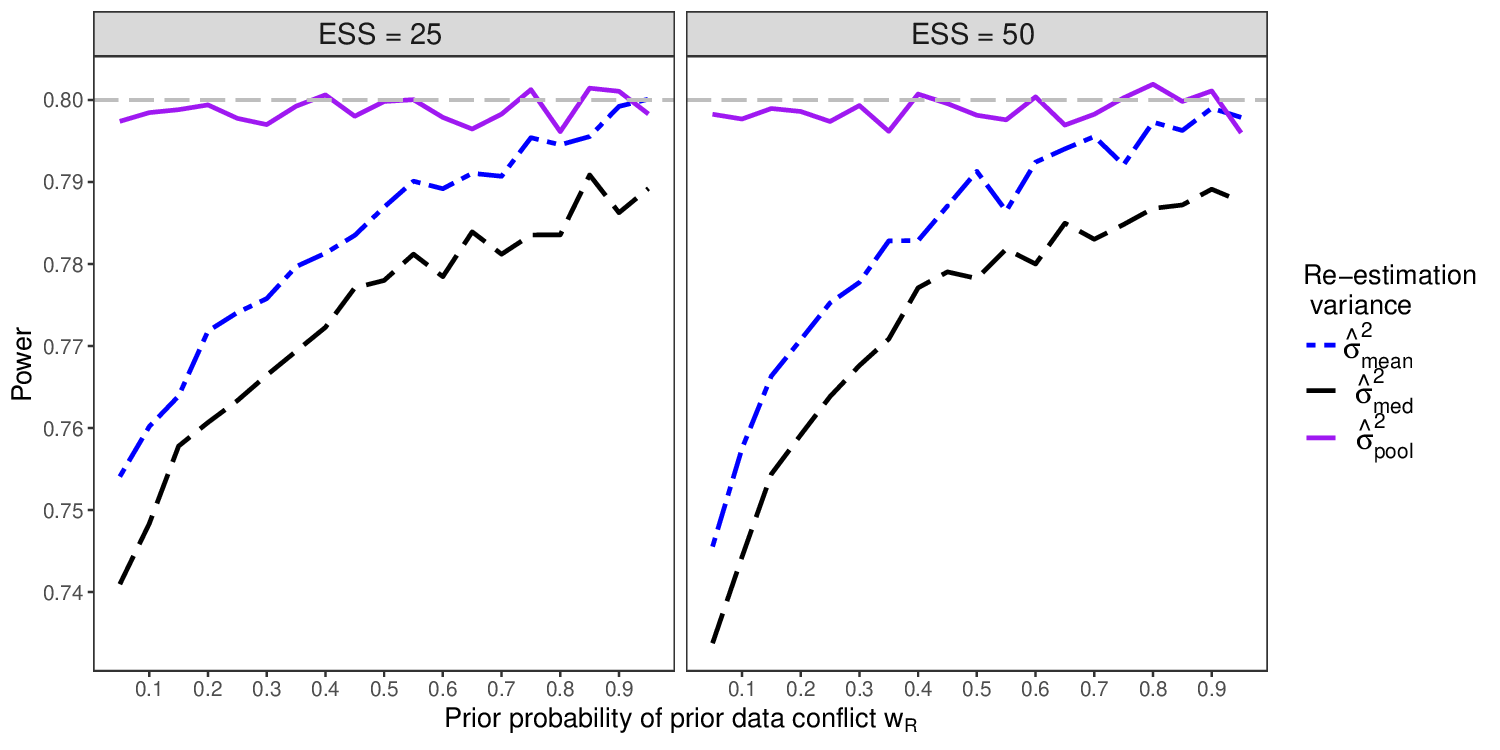}
  \caption{Power of the different sample size re-estimation procedures when the prior is robustified and a prior-data conflict is present. The sample size of the internal pilot study is $n_1=60$. The horizontal grey line depicts the nominal power of $80\%$.}
 \label{fig:SceRobustPower}
\end{figure}

\begin{figure}[htb]
 \centering
 \includegraphics[width=1\linewidth]{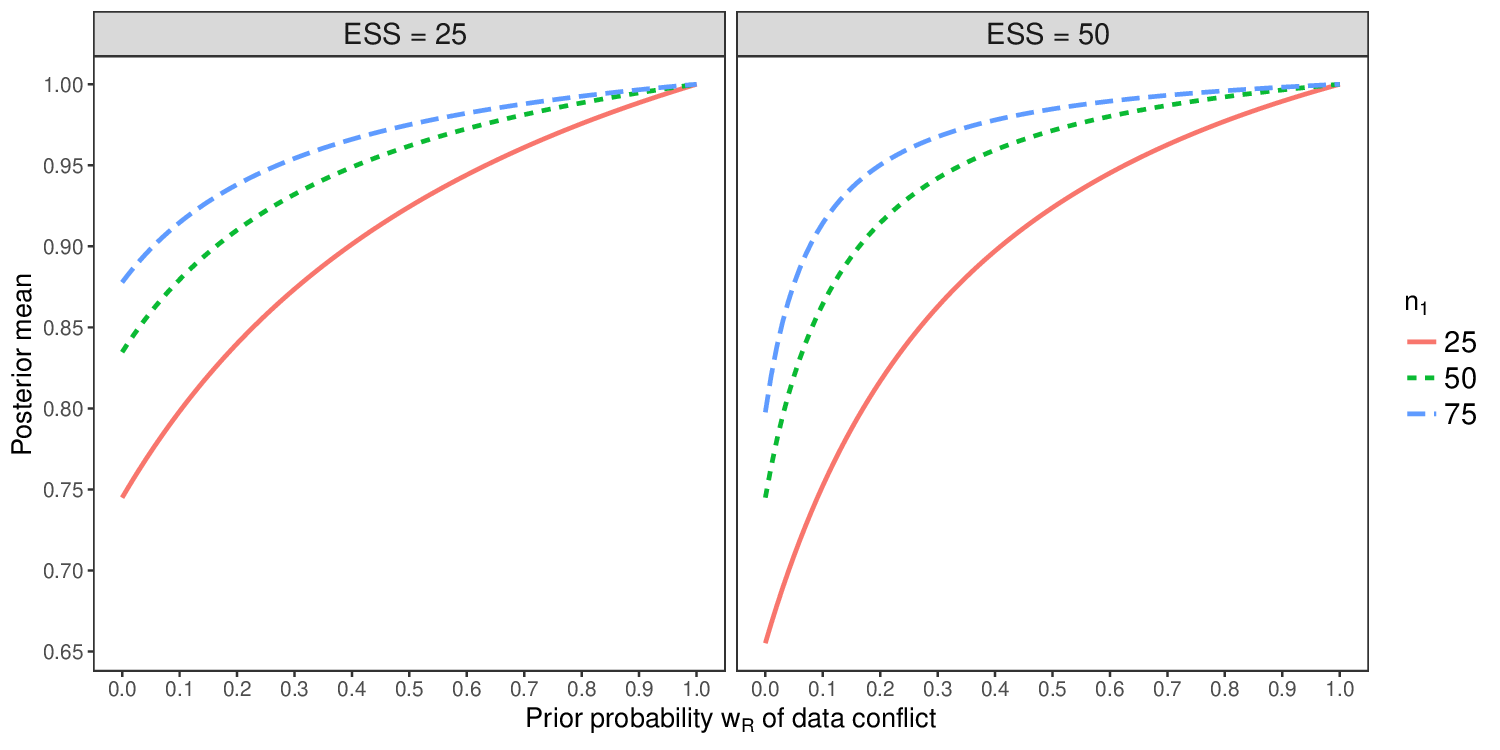}
  \caption{Posterior mean of the variance when the pooled variance estimator from the internal pilot study is equal to one, $\hat{\sigma}_{1,pool}^{2}=1$.}
 \label{fig:Postmean}
\end{figure}

The most important operating characteristic of a sample size re-estimation procedure is whether the test at the end of the study meets the target power under the condition that the type I error rate is controlled under the null hypothesis. 
Further operating characteristics are the distribution of the final sample size and the type I error rate.
In the following we assess the performance of Student's \textit{t}-test for the hypothesis $H_0$ at the end of the study after the sample size has been altered mid-study using the re-estimation procedure which incorporates prior information. 
The performance assessment is conducted by means of a Monte Carlo simulation study.
The simulation study is split into three parts. 
We start by considering an ideal setting in which the prior information is correct, in other words no prior-data conflict exists. 
This is simulated by choosing the expected value of the prior identical to the true variance $\sigma^2$ of the clinical trial.
Afterwards, we study the performance of the sample size re-estimation procedure for a prior-data conflict, which is simulated by choosing the expected value of the prior distribution considerably different to the true variance $\sigma^2$ of the clinical trial.
We conclude with scenarios which include a prior-data conflict but in which the prior distribution was also robustified.
Particular emphasis will be placed on the comparison of the sample size re-estimation procedure incorporating prior information with the sample size re-estimation approach based on the pooled variance.
Several methods for determining the sample size given a prior distribution were discussed in the previous section. 
During the presentation of the results of the simulation study we restrict ourselves to sample sizes re-estimated based on the Bayes estimators $\hat{\sigma}^{2}_{mean}$ and $\hat{\sigma}^{2}_{med}$ and do not include sample size re-estimation based on the unconditional power concerning $\sigma^2$ or on quantiles of the posterior distribution.
The performance is qualitatively the same with respect to the effects of the prior effective sample size, the internal pilot study, and the prior-data conflict and therefore not presented here.
Throughout this section, we select a Gamma distribution, not a mixture of Gamma distributions, as the MAP prior for the precision $\omega$. 
This simplified setting is already sufficient to highlight the main characteristics of the sample size re-estimation procedure incorporating prior information while not artificially increasing the number of parameters.
Moreover, the effective sample size can be illustrated easily for a Gamma distribution.
The prior effective sample size for a Gamma distribution with shape parameter $a$ and rate parameter $b$ is given by $ESS=2a$. 
Table \ref{table:Scenarios} lists the parameters considered in the simulation study.
\begin{table}[ht]
\caption{Scenarios for the Monte Carlo simulation study.}
\label{table:Scenarios}
\begin{center}
\begin{tabular}{l*{1} {c}{c}}
\toprule
\textbf{Parameter} & \textbf{Value}  \\ \midrule
One-sided significance level $\alpha$  & $0.025$ \\
Target power $1-\beta$  & 0.8 \\
Margin $\delta^{*}$ in the alternative & 0.5 \\
True variance $\sigma^2$ & 1 \\
Internal pilot study size $n_1$ & $10, 20, \ldots, 100$ \\
Sample size ratio $k$ & 1 \\
Expected value of prior $p_{\sigma^2}(\cdot)$ (no prior-data conflict) & 1 \\
Expected value of prior $p_{\sigma^2}(\cdot)$ (prior-data conflict) & 0.49\\
Effective sample size $ESS$ of prior $p_{\sigma^2}(\cdot)$ & $6, 25, 50$ \\
\bottomrule
\end{tabular}
\end{center}
\end{table}
As listed in Table \ref{table:Scenarios}, the true variance of the normally distributed data is $\sigma^2=1$. 
We assume a mean difference in the alternative of $\delta^{*}=0.5$.
For $\sigma^2=1$ and $\delta^{*}=0.5$, in a fixed design a sample size of $n=128$ would be required to obtain a power of $1-\beta=0.8$. 
However, it is important to emphasize that in practice the true variance and  the respective true fixed design sample size are not known.
Therefore, by varying the internal pilot study over a wide range we include the cases in the simulation study in which the internal pilot study sample size is specified based on a variance deviating from the true variance.
Moreover, as Table \ref{table:Scenarios} highlights, during the simulation study particular emphasis is put on the performance of the sample size re-estimation procedure when the internal pilot study sample size $n_1$ and the effective sample size change.
The shape parameter $a$ and rate parameter $b$ of the prior density $p_{\sigma^2}(\cdot)$ are chosen based on the effective sample size $ESS$ and the expected value of the prior distribution. 
The shape parameter is half of the effective sample size: $a=ESS/2$.
Let $\sigma^2_{mean}$ denote the expected value of the prior distribution of the variance.
Since the prior of the variance has an inverse Gamma distribution, the expected value is given by  $\sigma_{mean}^2 = b/(a-1)$.
Thus, the rate parameter is determined to be $b=\sigma_{mean}^2 ( ESS/2  - 1)$.
The expected value $\sigma^2_{mean}$ of the prior distribution is set to one to model the settings with no prior-data conflict.
On the other hand, to model a prior-data conflict, the expected value of the prior distribution is set to $\sigma^2_{mean}=0.49$.
Each simulated power in this section is based on $50\,000$ Monte Carlo replications which corresponds to a simulation error of less than $0.0018$ for a simulated power of $0.8$.
In the following the results of the Monte Carlo simulation study are presented for the scenarios without prior-data conflict listed in Table \ref{table:Scenarios}.
Figure \ref{fig:Sce1power} plots the power of Student's \textit{t}-test against the internal pilot study sample size $n_1$ for the different sample size re-estimation procedures.
Figure \ref{fig:Sce1power} shows, if no prior-data conflict is present, incorporating external data into the sample size re-estimation results in a power closer to the nominal level compared to the traditional sample size re-estimation approach, except for a small prior effective sample size of $ESS=6$. 
A sample size re-estimation based on the posterior mean $\hat{\sigma}_{1,mean}^{2}$ leads to better results than the re-estimation based on the posterior median $\hat{\sigma}_{1,med}^{2}$.
This is due to the equality of the prior mean and the true variance as well as the fact that the prior median is slightly smaller than the prior mean for the considered scenarios.
Moreover, the larger the prior effective sample size, the smaller is the  internal pilot study sample size required to reach to the nominal power level. 
In Figure \ref{fig:Sce1ss} the distribution of the final sample size $\hat{n}_{final}$, depicted by the median and the range between the $10\%$ and $90\%$ percentiles, is compared between the traditional sample size re-estimation procedure and the procedure incorporating prior information.
We do not present the results for the re-estimation procedure based on the posterior median since the results are qualitatively the same as for the procedure based on the posterior mean. 
Figure \ref{fig:Sce1ss} shows that the distribution of the final sample size is positively skewed for both sample size re-estimation procedures. 
Moreover, even incorporating prior information with a small prior effective sample size of $ESS=6$ into the sample size re-estimation results in a smaller variability of the final sample size.
The variability of the final sample size obtained with the sample size re-estimation procedure incorporating prior information decreases as the prior effective sample size increases. 
The difference in variability between the sample size re-estimation procedures with and without incorporated prior information decreases as the internal pilot study sample size increases since the larger the internal pilot study sample size, the smaller the effect of the prior on the posterior distribution. 
Additionally, we studied the type I error rate of the proposed sample size re-estimation procedure incorporating prior information on the variance and compared the results with the type I error rate of the sample size re-estimation procedure using the unblinded pooled sample variance.
The results are presented in Section 1 of the Supplementary Material.
The type I error rate of the proposed sample size re-estimation procedure incorporating prior information has a type I error inflation similar to the unblinded sample size re-estimation procedure based on the pooled sample variance for small prior effective sample sizes.
Moreover, the type I error rate of the proposed sample size re-estimation procedure incorporating prior information decreases and converges against the nominal level as the prior effective sample size increases.
In conclusion, incorporating prior information into the sample size re-estimation is beneficial when no prior-data conflict is present. \\ \indent
Next, we study how a prior-data conflict affects the performance of the sample size re-estimation procedure when the prior-data is incorporated into the sample size re-estimation. 
Thereto, we still consider a true variance of one, $\sigma^2=1$, but now the parameters of the prior density $p_{\sigma^2}(\cdot)$ are chosen such that the prior distribution has an expected value of $\sigma^2_{mean}=0.49$. 
The results of the corresponding Monte Carlo simulation study are presented in Figure \ref{fig:Sec2power}.
Figure \ref{fig:Sec2power} shows that sample size re-estimation incorporating a prior which conflicts with the actual data results in underpowered trials if the prior distribution has a mean or median smaller than the true variance.
Even for a small prior effective sample size of $ESS=15$, large internal pilot studies cannot resolve the prior-data conflict.
The underpowering increases as the effective sample size increases but decreases as the internal pilot study sample size increases. 
If the prior distribution would have a mean or median larger than the true variance, the sample size re-estimation incorporating prior information would overpower the clinical trial.
Thus, the sample size re-estimation procedure incorporating prior information using the MAP prior does not meet the basic performance requirement when the prior-data is in conflict with the variance of the ongoing clinical trial. \\ \indent
For the last part of this simulation study, we robustify the MAP prior as introduced above and assess whether the corresponding sample size re-estimation procedure incorporating prior information is robust against prior-data conflicts.
We assume that the MAP prior $p_{\sigma^2}(\cdot)$ on the variance has an expected value of $\sigma_{mean}^{2}=0.49$ while the true variance is $\sigma^2=1$. 
For the robustification we mix the MAP prior $p_{\sigma^2}(\cdot)$ with a vague prior $p_{V}(\cdot)$ which follows an inverse Gamma distribution with shape parameter $a=2$ and rate parameter $b=1$. 
The vague prior has a prior effective sample size of $ESS=4$ and its expected value is equal to one, i.e. the same as the true variance. 
Thus, this choice of vague prior corresponds to an idealized setting which aims to emphasize what can be achieved concerning power control by robustifying the MAP prior.
The robustified MAP prior of the precision is given by
\begin{align}
\label{eq:robustMix}
w_{R}\, \operatorname{Gamma}(2, 1) + (1-w_{R})\, \operatorname{Gamma}\left(ESS/2, 0.49(ESS/2-1) \right).
\end{align}
Moreover, the prior probability of a prior-data conflict, $w_{R}$, is varied within the simulation study between $0.05$ and $0.95$. 
A small value of $w_{R}$ corresponds to a small prior probability of a prior-data conflict.
Here, we focus on the results for an internal pilot study size of $n_{1}=60$ which corresponds to about half the sample size required in the fixed sample design. 
The effective sample size of the informative part of the robustified prior, $p_{\sigma^2}(\cdot)$, is set to $ESS=25, 50$. 
The results of the power simulation are shown in Figure \ref{fig:SceRobustPower}.
More detailed simulation results in which the internal pilot study size is varied too and more effective sample sizes are considered are shown in the Supplementary Material. 
Figure \ref{fig:SceRobustPower} shows that the larger the prior probability of a prior-data conflict, the less does the prior-data conflict affect the power of the final analysis after sample size re-estimation. 
However, Figure \ref{fig:SceRobustPower} also shows that the prior probability $w_{R}$ for a data conflict has to be very close to one in the case of a prior-data conflict to mitigate the effects of a prior-data conflict on the power.
Moreover, an increase of the prior probability $w_{R}$ also reduces the benefits of incorporating prior information into the sample size re-estimation when no prior-data conflict is present as the results shown in the Supplementary Material highlight.\\ \indent
Robustifying the MAP prior does not result in the desired power of the sample size re-estimation procedure for the case of a prior-data conflict because the information from the internal pilot study cannot appropriately discount the prior information.
We will illustrate the inability of the internal pilot study to discount incorrect prior information further in the following.
Thereto, we assume an observed pooled variance of one from the internal pilot study, i.e. $\hat{\sigma}_{1,pool}^{2}=1$. 
The prior information on the variance is the same as the mixture in Formula \eqref{eq:robustMix}.
In Figure \ref{fig:Postmean} the posterior mean of the variance is plotted against the prior probability $w_{R}$ of a prior-data conflict for the prior effective sample sizes $ESS=25, 50$ as well as the internal pilot study sizes $n_1=25, 50, 75$.
Figure \ref{fig:Postmean} shows only a large prior probability $w_{R}$ of a data conflict results in discounted prior information. 
The larger the internal pilot study sample size, the more the prior information is discounted for a fixed prior probability $w_{R}$.
Moreover, a larger prior effective sample size can also be beneficial in detecting a prior-data conflict.
However, for the considered practically relevant effective sample sizes and internal pilot study sizes the prior information cannot be discounted enough to not reduce the power of the sample size re-estimation procedure incorporating prior information in the case of a prior-data conflict. 
As for the case of no prior-data conflict, we also studied the performance of the proposed sample size re-estimation procedure concerning the type I error rate when a prior-data conflict is present and the MAP prior is robustified.
As highlighted in Section 1 of the Supplementary Material, incorporating robustified prior information into the sample size re-estimation results in a type I error inflation similar to the case of unblinded sample size re-estimation based on the pooled variance estimator.\\ \indent
Concluding, incorporating prior information into the sample size re-estimation results in a power closer to the target power compared to the sample size re-estimation based on the pooled variance estimator in the case of no prior-data conflict. 
Moreover, the variability of the final sample size also decreases when correct prior information are incorporated into the sample size re-estimation.
However, when the prior information conflicts with the data from the internal pilot study, incorporating prior information into the sample size re-estimation leads to under- or overpowered clinical trials. 
The adverse influence of a prior-data conflict can be limited, but not corrected, by robustifying the prior information which, however, reduces the benefit of incorporating prior information into the sample size re-estimation in the case of no prior-data conflict.
\section{Blinded sample size re-estimation}
\label{sec:blinded}
In this section we discuss various ways of incorporating prior information into the blinded sample size re-estimation and we outline why the resulting procedures face the same obstacles as the unblinded procedures in the case of a prior-data conflict. 
The one-sample variance estimator, which estimates the unknown variance by the sample variance of the blinded data, is the recommended method for blinded sample size re-estimation in two-arm trials with continuous data. \cite{friede2013blinded}
In contrast to sample size re-estimation based on the unbiased pooled sample variance, sample size re-estimation based on one-sample variance estimator meets the target power due to overestimating the outcome variance.
The relationship between the one-sample variance estimator $\hat{\sigma}^{2}_{OS}$  and the pooled sample variance can be illustrated by the following equation\cite{Gould1992sample}
\begin{align*}
\hat{\sigma}^{2}_{OS} = \frac{n_{1}-2}{n_{1}-1}\hat{\sigma}^2_{pool} + \frac{n_{1T}n_{1C}}{n_{1}(n_{1}-1)}\left(\bar{X}_{1T} - \bar{X}_{1C}\right)^2.
\end{align*}
Since simply replacing the pooled variance estimator with the one-sample variance estimator resulted in the desired properties of the sample size re-estimation procedure without prior information, it seems natural to proceed similar when incorporating prior information into the sample size re-estimation.
Thus, when updating the prior information on the variance the pooled sample variance could be replaced by the one-sample variance estimator. 
Doing so increases the power of the sample size re-estimation procedure incorporating prior information. 
In simulations not presented here we observed that in the case of a prior-data conflict the power increase is small compared to the deviations from the target power and that the prior information are not discounted enough to actually obtain a suitable procedure for incorporating prior information into the sample size re-estimation in the case of a prior-data conflict.\\ \indent
An alternative method for blinded sample size re-estimation in two-arm trials with a randomized block design is based on the Xing-Ganju variance estimator which estimates the outcome variance blinded based on the block sums. \cite{xing2005method}
Sample size re-estimation incorporating prior information can be extended to not require unblinding by focusing on the likelihood of the block sums when updating the prior information.
In more detail, let $b$ be the number of randomized blocks and let $m$ be the number of observations within each block. 
For the sake of simplicity, we assume equal allocation between treatment arms within each block.
Conditioned on the means $\mu_{T}$ and $\mu_{C}$ and the variance $\sigma^2$, the block sums $T_{i}$, $i=1,\ldots, b$, are normally distributed,
\begin{align*}
T_{i}|\mu_{T}, \mu_{C}, \sigma^2 \sim \mathcal{N}\left(0.5m(\mu_{T} + \mu_{C}), m\sigma^2\right).
\end{align*}
Thus, $S_{i}=T_{i}/\sqrt{m}$ follows conditionally a normal distribution with mean $0.5\sqrt{m}(\mu_{T} + \mu_{C})$ and variance $\sigma^2$.
In Section \ref{sec:SSR} we considered improper uniform priors for the means.
Under this assumption, the prior for the sum of means is also an improper uniform prior.
The formulas for updating a Gamma mixture prior for the precision can be easily adapted from Section \ref{sec:SSR} by using the Xing-Ganju estimator 
\begin{align*}
\hat{\sigma}_{XG}^{2}=\frac{1}{b-1}\sum_{i=1}^{b}(S_{i} - \bar{S}_{\cdot})^2
\end{align*}
instead of the pooled variance estimator and noting that the degrees of freedom are $b-1$ instead of $n_1-2$.
That being said, the similarity of the models also implies that the performance of the resulting sample size re-estimation procedures incorporating prior information will be very similar to the case of the unblinded sample size re-estimation with prior information. 
The lower number of degrees of freedom of the Xing-Ganju variance estimator further reduces the ability to discount prior information in the case of a prior-data conflict, too. 
Thus, blinded sample size re-estimation incorporating prior information based on the Xing-Ganju variance estimator will not achieve the target power in the case of a prior-data conflict.\\ \indent
The blinded data from the internal pilot study follow a mixture of two normal distributions. 
Therefore, it has been studied to estimate the variance for the blinded sample size re-estimation using the expectation-maximization (EM) algorithm. \cite{Gould1992sample, friede2002inappropriateness, waksman2007assessment}
However, the variance estimator from the EM algorithm is biased downwards resulting in a sample size re-estimation procedure underpowering clinical trials.
When updating the prior information on the variance after the internal pilot study using the likelihood of a mixture of normal distributions, 
similar effects can be observed.
The posterior distribution for the variance has a mean and median smaller than the pooled variance. 
Thus, considering those location parameters in the sample size re-estimation does also result in underpowering the clinical trial in the case of a prior-data conflict. \\ \indent
Concluding, prior information can easily be incorporated into the blinded sample size re-estimation.
However, similar to the unblinded sample size re-estimation incorporating prior information, in the case of a prior-data conflict the prior information cannot be discounted enough to ensure that the resulting sample size re-estimation procedure results in a clinical trial meeting the target power.
\section{Discussion}
\label{sec:discussion}
In this manuscript we studied various methods for incorporating prior information on the nuisance parameter into the nuisance parameter based sample size re-estimation.
The prior information was given as an MAP prior which is obtained through a meta-analysis of variances from historical clinical trials using a Bayesian hierarchical model. 
In the case of no prior-data conflict, that is when the prior mean corresponds to the sample variance of the ongoing clinical trial, incorporating prior information into the sample size re-estimation decreases the variability of the final sample size compared to the unblinded sample size re-estimation based on the pooled sample variance. 
Moreover, in contrast to the unblinded nuisance parameter based sample size re-estimation procedure, the sample size re-estimation procedure incorporating prior information on the nuisance parameter meets the nominal power when the prior information is correct. 
However, the sample size re-estimation approach incorporating prior information is not robust concerning prior-data conflicts in which case it leads to under- or overpowered clinical trials.
Robustifying the MAP prior improves the performance of the sample size re-estimation approach incorporating prior information, however, not to the point where the clinical trials are not under- or overpowered. 
This is due to the internal pilot studies not containing enough information to sufficiently discount the misspecified prior information.
The performance of the nuisance parameter based sample size re-estimation procedure incorporating prior information was primarily studied for unblinded data even though the use of blinded data is generally recommended when the sample size is adjusted based on a nuisance parameter estimate.
That being said, we proposed several approaches to incorporate prior information into the blinded sample size re-estimation and we exemplified that the resulting blinded sample size re-estimation procedures incorporating prior information do not meet the target power either when a prior-data conflict is present.
Concluding, incorporating prior information on the nuisance parameter into the nuisance parameter based sample size re-estimation can be beneficial in the sense that it reduces the variability of the final sample size compared to the sample size re-estimation without prior information.
However, incorporating prior information on the nuisance parameter into the sample size re-estimation also bears the risk of resulting in an under- or overpowered clinical trial when a prior-data conflict is present. 
The severity of missing the target power in the case of a prior-data conflict depends on the size of the prior-data conflict, on the prior effective sample size, and on the internal pilot study sample size.
Thus, the decision whether to incorporate prior information on the nuisance parameter into the sample size re-estimation can only be made on a case-to-case basis after carefully weighting the risks with the potential benefits.
It is also worth emphasizing that for the studied clinical trial setting the traditional nuisance parameter based sample size re-estimation using the one-sample variance estimator does not bear the risk of missing the target power.
\\ \indent
The sample size re-estimation procedure incorporating prior information inflates the type I error rate in a similar magnitude as the unblinded sample size re-estimation procedure based on the pooled sample variance. 
However, strategies for controlling the type I error inflation have been proposed in the literature. \cite{Friede2010b, friede2010}\\ \indent
In this manuscript we robustified the MAP prior from historical data by mixing it with a vague prior. 
The prior probability $w_R$ of a prior-data conflict is prespecified, i.e. it is not a random variable and does not depend on the data which is used to update the prior distribution, as it was proposed by Schmidli et al.\cite{schmidli2016meta}
It should be noted that the weight $w_R$ will  be different in the posterior distribution. For example, if there is a prior-data conflict, this weight will be lower. \\ \indent
Hartley studied blinded sample size re-estimation for normal data and concluded that incorporating prior information into the blinded sample size re-estimation is generally recommended. \cite{hartley2012adaptive}
It is worth highlighting the differences between Hartley's publication and our manuscript since we draw a different conclusion.
In Hartley's publication, prior information on both the variance and the effect are incorporated into the sample size re-estimation which selects the final sample size based on a predictive power.
In contrast, we did not consider uncertainty in the effect size and we studied the setting where the final sample size is determined based on the sample size formula for Student's \textit{t}-test as it is common in nuisance parameter based sample size re-estimation for a frequentist analysis at the end of the clinical trial.
Moreover, in this manuscript we focused on the overarching approach of both summarizing the prior information based on the MAP approach and incorporating the gathered prior information into the nuisance parameter based sample size re-estimation.
We also extensively studied the case of prior-data conflicts which eventually lead to our conclusion that the decision to incorporate prior information on the nuisance parameter into the sample size re-estimation has to be made on a case-to-case basis.\\ \indent
The focus of this manuscript was nuisance parameter based sample size re-estimation in the case of a two-arm parallel group superiority trial with normally distributed outcomes.
Nuisance parameter based sample size re-estimation has been proposed for other designs such as designs with more than two treatment arms or designs with count data. \cite{mutze2016blinded, friede2010blinded}
The performance of the nuisance parameter based sample size re-estimation procedures incorporating prior information in these designs might differ due to larger internal pilot studies or endpoints for which prior-data conflicts can be detected easier. \\ \indent 
We focused on nuisance parameter based the sample size re-estimation.
Others have proposed to adjust the sample size of a clinical trial based on treatment effect estimates. \cite{lehmacher1999adaptive}
Additionally, the use of Bayesian approaches has already be advocated for monitoring group sequential designs.\cite{gsponer2014practical}	
In future research, incorporating prior information about the effect into the sample size re-estimation should be studied.
Moreover, while we presented a re-estimation approach which adjusts the sample size by plugging in a Bayes point estimator into the sample size formula, other rules for selecting the final sample size could be chosen. 
For instance, decision theoretic methods could be used to derive possibly better rules for the sample size re-estimation. \cite{stallard1998sample} \\ \indent
The simulation and calculations presented in this manuscript were performed using the R language. \cite{rcore}
To reproduce the results presented in this manuscript, the respective code has been made available through the R package \textit{varmap}.\cite{muetzegithub}

\section*{Acknowledgement}
Tobias M{\"u}tze is supported by the DZHK (German Centre for Cardiovascular Research) under grant GOE SI 2 UMG Information and Data Management.
The authors would like to thank two anonymous reviewers and an Associate Editor for their helpful and constructive comments.

\end{document}